**TITLE**

**Evidence of high-temperature exciton condensation in a two-dimensional semimetal**

**AUTHORS**

Qiang Gao[1]*, Yang-hao Chan[2,3]*, Yuzhe Wang[4], Haotian Zhang[1], Jinxu, Pu[1], Shengtao Cui[5], Yichen Yang[6], Zhengtai Liu[6], Dawei Shen[6], Zhe Sun[5], Juan Jiang[4], Tai C. Chiang[7,8#], Peng Chen[1#]

[1]Key Laboratory of Artificial Structures and Quantum Control (Ministry of Education), Shenyang National Laboratory for Materials Science, Shanghai Center for Complex Physics, School of Physics and Astronomy, Shanghai Jiao Tong University, Shanghai 200240, China.

[2]Institute of Atomic and Molecular Sciences, Academia Sinica, Taipei 10617, Taiwan.

[3]Physics Division, National Center for Theoretical Sciences, Taipei 10617, Taiwan.

[4]School of Future Technology, University of Science and Technology of China, Hefei, Anhui 230026, China

[5]National Synchrotron Radiation Laboratory, University of Science and Technology of China, Hefei, Anhui 230026, China.

[6]State Key Laboratory of Functional Materials for Informatics, Shanghai Institute of Microsystem and Information Technology (SIMIT), Chinese Academy of Sciences, Shanghai 200050, China.

[7]Department of Physics, University of Illinois at Urbana-Champaign, 1110 West Green Street, Urbana, Illinois 61801-3080, USA.

[8]Frederick Seitz Materials Research Laboratory, University of Illinois at Urbana Champaign, 104 South Goodwin Avenue, Urbana, Illinois 61801-2902, USA.

*These authors contributed equally to this work.

#e-mail: tcchiang@illinois.edu; pchen229@sjtu.edu.cn

## ABSTRACT

**Electrons and holes can spontaneously form excitons and condense in a semimetal or semiconductor, as predicted decades ago. This type of Bose condensation can happen at much higher temperatures in comparison with dilute atomic gases. Two-dimensional (2D) materials with reduced Coulomb screening around the Fermi level are promising for realizing such a system. Here we report a change of the band structure accompanied by a phase transition at about 180 K in single-layer ZrTe₂ based on angle-resolved photoemission spectroscopy (ARPES) measurements. Below the transition temperature, gap opening and development of an ultra-flat band top around the zone center are observed. This gap and the phase transition are rapidly suppressed with extra carrier densities introduced by adding more layers or dopants on the surface. The results suggest the formation of an excitonic insulating ground state in single-layer ZrTe₂, and the findings are rationalized by first-principles calculations and a self-consistent mean-field theory. Our study provides evidence for exciton condensation in a 2D semimetal and demonstrates strong dimensionality effects on the formation of intrinsic bound electron-hole pairs in solids.**

## INTRODUCTION

Excitons in solids, bound states between an electron and a hole, were predicted to condense under appropriate conditions[1-8]. Because of their small effective masses, such condensation can occur at high temperatures[9,10], which is of great interest in modern condensed matter research. Although the condensed phases of excitons have been realized in cases involving optical pumping, quantum Hall states in a magnetic field, and electrical generation in atomic double-layer heterostructures[9-14], very few examples were identified in which the long-range excitonic order is spontaneously formed at low temperatures under equilibrium conditions. Realization of such a system is difficult because it needs a delicate balance among several factors: the size of the band gap, the binding energy of the electron-hole pairs, and the screening strength[5,7]. Promising candidates so far include TmSe$_{0.45}$Te$_{0.55}$ under pressure, Ta$_2$NiSe$_5$, 1$T$-TiSe$_2$, and other proposed candidates[15-29]. It is still under intense debate whether the phase transition in Ta$_2$NiSe$_5$ is driven by an exciton condensation or the lattice distortion from change of the structure,[22,23,27,28] and various competing interpretations have been offered for bulk TiSe$_2$[18,20]. Thus far, experimental observation of a spontaneous exciton condensation remains inconclusive.

Exciton condensation in materials has been predicted to lead to a new insulating ground state, an excitonic insulator[4-8]. General features related to the phase transition include the mixing of conduction and valence bands and gap opening around the Fermi level ($E_F$). The band structure is renormalized and exhibits a characteristic flat valance band top (Fig. 1)[29]; these signatures should be readily captured by ARPES. Transition metal dichalcogenides with vanishing indirect band gap or band overlap between the valence and conduction bands are promising candidates for developing 2D excitonic insulators[4-7]. These layered materials can be easily prepared as single layers, and single-layer TiSe$_2$ appears to be a natural choice. However, the semimetal-insulator transition and a flat valence band top have not been observed[30]. Another strong candidate is ZrTe$_2$; in bulk form, it is a metal/semimetal with a negative band gap of about -0.5 eV[31,32]. The Coulomb interaction between electrons and holes is strongly screened by the many carriers around the metallic Fermi surfaces, and no exciton condensation is expected or observed. This unfavorable condition is, however, completely reversed for the single layer, for which the gap is almost zero and electronic screening is largely suppressed in the 2D limit[26,33].

**RESULTS**

**Electronic band structure of single-layer ZrTe$_2$**

Here we study the band structures of single-layer ZrTe$_2$ and present experimental evidence of high-temperature exciton condensation. The structure of single-layer ZrTe$_2$ (Fig. 1a) in the normal phase consists of a triangular planar net of Zr atoms sandwiched between two Te atomic layers. Sharp reflection-high-energy-electron diffraction (RHEED) patterns (Fig. 1c) reveal a high-quality and well-ordered single layer of ZrTe$_2$ grown by molecular beam epitaxy (MBE) on a bilayer-graphene-terminated SiC substrate. Angle-integrated core level scans (Fig. 1d) show the characteristic peaks of Zr and Te for the 1$T$ phase[31]. Fig. 1f shows ARPES maps taken with $s$ polarized light along the $\overline{\Gamma M}$ direction. Two hole-like valence bands are seen centered at the $\overline{\Gamma}$ point. A previous study of single-layer ZrTe$_2$ grown on InAs (111) shows a similar band structure, but with an upward shift of the Fermi level. The difference suggests a charge transfer from the InAs substrate[32]. For our sample at 300 K, the conduction band edge at the $\overline{M}$ point reaches just around the Fermi level. With the valence band edge just slightly above the Fermi level, the system is metallic. The indirect band overlap around the Fermi level is only about -0.06 eV (Supplementary Fig. 2), much smaller than the bulk case. A small band overlap implicates a small carrier density around the Fermi level, which is favorable for the formation of excitons.

Upon cooling to low temperatures, the topmost valence band develops a dispersionless (flat) top around the zone center in a range of ~ ±0.072 Å$^{-1}$(Fig. 1f). The downward shift of the band top results in a gap of 97 meV from the Fermi level at the zone center at 10 K. This strong band renormalization is consistent with the excitonic insulator scenario that excitons consist of holes in the valence band and electrons in the conduction band condense naturally in single-layer ZrTe$_2$. The flat band dispersions correspond to a zero group velocity for the electrons and holes in the condensed state. As seen in Fig. 2a, the flat valence band top at 10 K becomes much weaker in spectra taken with $p$ polarized light. Another relevant feature of exciton formation is the hybridization of valence and conduction bands, which gives rise to strong folded valence bands, as observed around the $\overline{\text{M}}$ point at 10 K (Fig. 2a).

**Excitonic insulator model**

These observations can be understood in terms of an excitonic phase based on a BCS-like model[6]. Calculated spectral functions for the valence and conduction bands along $\overline{\Gamma\text{M}}$ are shown in Fig. 2b. The results for the normal phase reproduce the essential features of the ARPES data including the spin-orbit splitting of the two valence bands. For the condensed phase, a mean-field self-consistent excitonic order parameter $\Delta_X$ is included in the calculation to account for, to first order, the hybridization of conduction and valence bands. The main effects are to flatten the valence band top and to open a gap of 74 meV, both in agreement with the ARPES results. The flat valence band top around the zone center gets folded to the $\overline{\text{M}}$ point with considerable spectral weight transfer, which indicates a strong electronic nature of the phase transition[19].

**Temperature dependence of the energy gap**

Systematic scans of the bands along $\overline{\Gamma\text{M}}$ (Fig. 3a) as a function of temperature reveal details of the development of the flat band feature, opening of the gap, and the intensity variation of the folded bands in connection with the semimetal-insulator transition. The gap formation is illustrated by the symmetrized ARPES maps in Fig. 3b. The symmetrized energy distribution curves (EDCs) at the zone center at various temperatures are shown in Fig. 3c. At high temperatures, a single peak at the Fermi level at $E_F$ indicates that the valence band top crosses over $E_F$. At lower temperature, this single peak bifurcates, indicating the opening of a gap. The energy gap, determined from the difference of the valence band top and the conduction band bottom, is 75 meV at 10 K. The square of the extracted energy gap is plotted as a function of temperature in Fig. 3d. Fitting to a BCS mean-field equation[34] (red curve in Fig. 3d) yields a transition temperature of $T_C = 180 \pm 6$ K.

**Charge density wave (CDW) order**

The nature of the phase transition is complicated by the possibility of a Peierls-type CDW transition driven by lattice instability, as opposed to an excitonic transition driven by electron-hole interaction. We performed first-principles calculations for a $(2 \times 2)$ superstructure to check the effects of lattice instability (Supplementary Fig. 6). The calculated total energy per chemical unit is lower by just 0.4 meV for the $(2 \times 2)$ phase relative to the normal phase for single-layer $ZrTe_2$. This is much smaller than the corresponding energy lowering of 5 meV for the $(2 \times 2)$ CDW in single-layer $TiSe_2$ [30]. The anharmonic well potential indicates intrinsic lattice instability in $TiSe_2$, whereas the effect is not obvious in $ZrTe_2$ (Supplementary Fig. 8b). Calculated atomic displacements in the CDW phase show $\delta Zr$ is 0.7% of the lattice constant (0.028 Å) and $\delta Te$ is 0.3% of the lattice constant. As a comparison, the amplitude of $\delta Ti$ is 2.5% of the lattice constant (0.088 Å) in the single layer $TiSe_2$ in the CDW phase[30,35]. Although an imaginary phonon mode is found for single-layer $ZrTe_2$ at the $\bar{M}$ point, the PBE band structure shows no gap opening, indicating a weak lattice instability compared to the other known CDW cases (Supplementary Figs. 6 and 10).

To estimate how large the lattice distortion is necessary to open a gap of 75 meV observed by ARPES, we imposed various amounts of lattice distortion (represented by $\delta Zr$) in calculations and the obtained value is ~3.15% of the lattice constant, 4.5 times of the optimized CDW structure. Hubbard U = 6 eV were further included to account for the electron localization and the gap remains closed in the CDW phase where $\delta Zr = 0.7\%$ of the lattice constant. (Supplementary Fig.

8c, d). These findings indicate the insufficient ability of the lattice distortion to open a significant gap in this system, although it may contribute to the formation of the backfolded bands at the $\bar{M}$ points as in the case of single layer TiTe$_2$ ($\delta$Ti = 1.0% of the lattice constant)[36]. Effects of electron-hole hybridization can be better described using the HSE functional instead of the PBE in the calculation. Similar to the mean-field self-consistent method mentioned above, HSE calculations yield a gap of ~10 meV near the $\bar{\Gamma}$ point, indicating the necessity of the electron-hole interaction in this system. Of note is that DFT results predict a much larger overlap between the valence band top and the conduction band bottom than the experiment in the normal phase. The absolute value of the quantity predicted by the calculations may not be accurate due to the inherent limitations of the DFT method. However, the relative comparison between TiSe$_2$ and ZrTe$_2$ should remain valid.

The structural distortion in a single layer film can be revealed by low energy electron diffraction (LEED). The elastic scattering of the incident electrons with the crystal potential mostly comes from the core electrons and nuclei of crystals[37]. LEED patterns of single layer TiSe$_2$ show the 1/2 order diffraction spots corresponding to the (2 × 2) CDW phase at low temperatures (Supplementary Fig. 9a, b). However, no CDW diffraction spots were observed for single layer ZrTe$_2$ in the probed beam energy range (20-200 eV), indicating the small lattice distortion in the CDW phase. Consistent results were obtained for different batches of samples. As a first order approximation, the proportion of CDW spot intensity ($I_{CDW}$) to the corresponding Bragg spot intensity ($I_{Bragg}$) is proportional to the square of the atomic displacements ($\mu$) in the CDW phase[38]. With background intensity subtracted, $I_{CDW}/I_{Bragg}$ for single layer TiSe$_2$ is ~0.3. The noise level divided by the saturated intensity is used to establish the detection limit and the value is ~0.05, indicating $\mu_{ZrTe_2}$ should be less than 2/5 of $\mu_{TiSe_2}$ in average (Supplementary Fig.

9c-e), in consistent with the above prediction that $\mu_{ZrTe_2} \sim \mu_{TiSe_2}/3$. Evidently, our LEED results set the upper bound of the atomic displacement of single layer $ZrTe_2$ in the CDW phase and the phase transition in this material cannot be described in terms of the usual Peierls-type CDW.

**Dimensionality and carrier screening effect**

Since no exciton condensation occurs in bulk $ZrTe_2$, an interesting question is how such a transition evolves with the sample thickness from the 2D limit of a single layer to the bulk limit. ARPES spectra for $N$ = 1, 2, 2.5 trilayer (TL) $ZrTe_2$ taken with the unpolarized light from a He-discharge lamp along the $\overline{\Gamma M}$ direction are presented in Fig. 4 (the $N$ = 2.5 case represents a mixture of $N$ = 2 and 3). The measured dispersion relations at 300 K and the multiplication of bands as $N$ increases are in excellent agreement with theoretical results for the normal phase. All cases are metallic in the normal phase, with bands crossing the Fermi level. However, only the 1 TL sample shows gap opening and valence band folding at 10 K. Thus, the excitonic phase in the single layer is suppressed by adding just one TL. This suppression can be attributed to changes in the carrier density, which can be extracted from the Luttinger area of the Fermi surface[39]. The carrier density around the $\overline{M}$ point for a 2 TL film is ~4.8 x $10^{13}$/cm$^2$, which is 136% larger than the 1 TL case. The extra carrier density leads to a stronger screening between the electrons and holes, thus hampering the formation of the excitons. A very weak CDW order may still exist in multilayer films[40], but it does not lead to an excitonic phase.

To further explore the effects of carrier screening, we have employed *in situ* surface doping of K on the 1 TL sample (Supplementary Fig. 1a). With the sample at 10 K, increasing amounts of K doping cause the valence band top to move upward and poke through $E_F$ at a critical carrier density of 3.8 x $10^{13}$/cm$^2$. The folded valence bands also disappear at this critical doping level. The lower valence band, separated by spin-orbit splitting from the top valence band, also shifts up slightly upon initial K doping. However, further increasing the doping level beyond the critical value causes it to shift downward together with the conduction band. Thus, the overall movements of the bands cannot be described by a simple rigid shift of all bands, other effects like atom intercalation can renormalize the band structure[41]. An implication is that electron doping leads to the suppression of the bound electron-hole pairs; beyond the critical doping level, the gapped ground state is destroyed, and the system returns to a normal semimetallic state.

The total evidence based on both experiment and theory suggests that single-layer $ZrTe_2$ is a prototypical 2D excitonic matter. A strong interband electron-hole coherent coupling in 2D and poor electronic screening by a low carrier density around the Fermi level conspire to make this system a unique case to display the requisite flat bands for excitonic condensation. The results are relevant to the development of a comprehensive understanding of the physics of exciton condensation in solids. The high transition temperature in this case suggests a versatile platform for experimentation in quantum effects and phenomena based on exciton coherence.

## METHODS
### ARPES measurements

$ZrTe_2$ thin films were grown *in situ* in the integrated MBE/ARPES systems at the lab in Shanghai Jiao Tong University, beamlines 10.0.1 at Advanced Light Sources in Lawrence Berkeley National Laboratory, and beamline 13U at National Synchrotron Radiation Lab in Hefei. Samples were transferred via a vacuum suitcase to beamline 03U at Shanghai Synchrotron Radiation Facility[42]. Substrates of 4H-SiC were flash-annealed for multiple cycles to form a well-ordered bilayer graphene on the surface. Single-layer and multilayer $ZrTe_2$ were grown on top of the substrate by co-evaporating high purity Zr and Te from an electron-beam evaporator and a Knudsen effusion cell, respectively, while the substrate was maintained at 600 °C. The growth process and thickness of the films was monitored by RHEED, and the growth rate was set to 60 minutes per layer of $ZrTe_2$. Formation of the multilayer of $ZrTe_2$ is also evidenced by evolution of the band structure measured by ARPES. ARPES measurements were performed at a base pressure of ~5x10$^{-11}$ mbar with in-laboratory He discharge lamp (He-I 21.2 eV) and 16-100 eV photons at synchrotron using Scienta DA30 analyzers. Energy resolution is better than 15 meV and angular resolution is around 0.2°. Each sample's crystallographic orientation was precisely determined from the symmetry of constant-energy-contour ARPES maps.

**Computational details of density functional theory from first principles**

Calculations were performed using the Vienna ab initio package (VASP)[43,44] with the projector augmented wave method[45,46]. A plane-wave energy cut-off of 520 eV and a $24 \times 24 \times 1$ $k$-mesh were employed. The generalized gradient approximation (GGA) with the Perdew-Burke-Ernzerhof (PBE) functional[47] was used for structure optimization of the single-layer $ZrTe_2$. Freestanding films were modeled with a 16-Å vacuum gap between adjacent layers in the supercell. The fully optimized in-plane lattice constant for a single layer 1$T$-$ZrTe_2$ is $a = 3.974$ Å. Since PBE functional is known to underestimate band gap, HSE functional[48] was also employed to better capture electron-hole hybridizations. For the 2 TL and 3 TL, the PBEsol functional[49] was used to relax interlayer distance; the in-plane lattice constant was taken from the monolayer results. Spin-orbit couplings are included in the calculations. Phonon calculations were carried out using the supercell method as implemented in the Phonopy package[50]. The band unfolding was performed using the BandUP code[51,52].

**Self-consistent mean-field theory**

We study the excitonic insulator phase based on the mean-field calculation of a 11-band model. The noninteracting part of the model Hamiltonian was constructed from wannierized orbitals[53] of a first-principle HSE calculations. $H^0 = \sum \epsilon_{nk} c_{nk}^\dagger c_{nk}$. We consider interactions between two top most valence bands and two lowest conduction bands. The interaction part of the Hamiltonian reads

$$H^{int} = \frac{1}{N} \sum_{k,k',q} V(q) c_{ck}^\dagger c_{ck+q} c_{vk'}^\dagger c_{vk'-q}, \qquad (1)$$

where $N$ is the number of $k$-points in the Brillouin zone (BZ). We assume that the interaction does not depend on band indices and crystal momentum $k$ and $k'$ and takes a form of screened Coulomb interaction $V(q) = \frac{U \tanh \frac{\xi q}{2}}{\xi q/2}$[54], where $\xi$ is set to 25 nm and U is a tuning parameter. To solve the excitonic insulator state, we consider a set of order parameters of electron-hole coherence $\Delta_X(\mathbf{Q}) = \frac{1}{N} \sum_k \langle c_{ck}^\dagger c_{vk+Q} \rangle$ with wavevectors $\mathbf{Q} = \mathbf{b}_1/2, -\mathbf{b}_2/2$, and $(\mathbf{b}_2 - \mathbf{b}_1)/2$, where $\mathbf{b}_1$ and $\mathbf{b}_2$ are two in-plane reciprocal lattice vectors. These wavevectors correspond to the triple-q CDW states observed in the experiment[40].

We numerically solved the Hamiltonian in a folded BZ with a uniform sampling of 48x48 $k$-grid in the BZ. A $k$-point in the full BZ are labeled by a corresponding $k$-point in the folded BZ and a wave vector connecting them. We found that with a parameter $U = 65$, the result agrees well with the experiment. The self-consistent calculation is initialized with a nonzero order parameter then the mean-field Hamiltonian is diagonalized. From the eigenvectors we can compute the new order parameters and repeat the procedure until convergence. The spectral function is computed following Ref.[55]. We have

$$A(k, \omega) = \sum_n \sum_j \left| \varphi_{nk}^{j,q_0}(k) \right|^2 \delta(\omega - E_{nk}), \qquad (2)$$

where $E_{nk}$ and $\varphi_{nk}^{j,q_0}$ are the n-th eigenenergy and eigenfunctions of the mean-field Hamiltonian with momentum k, respectively, j is the index of the bare band and $q_0 = 0$ labels components from the folded BZ centered at $\bar{\Gamma}$ since the momentum k in the spectral functions is defined in the unfolded zone.

## DATA AVAILABILITY

General methods, experimental procedures, and characterizations are available within the article and the Supplementary Information. Other relevant data are available from the corresponding author upon reasonable request.

**ACKNOWLEDGMENTS**

We thank Prof. W. T. Zhang for helpful discussions. The work at Shanghai Jiao Tong University is supported by the Ministry of Science and Technology of China under Grant No. 2022YFA1402400 and No. 2021YFE0194100, the Science and Technology Commission of Shanghai Municipality under Grant No. 21JC1403000. Y. H. C. acknowledges support by the Ministry of Science and Technology, National Center for Theoretical Sciences (Grant No. 110-2124-M-002-012) and National Center for High-performance Computing in Taiwan. Part of this research is supported by ME2 project under Contract No. 11227902 from National Natural Science Foundation of China. J. J. acknowledges support from the National Natural Science Foundation of China (Grant No. 12174362). T. C. C. acknowledges the support from U.S. Department of Energy, Office of Science, Office of Basic Energy Sciences, Division of Materials Science and Engineering, under Grant No. DE-FG02-07ER46383. P. C. thanks the sponsorship from Yangyang Development Fund.

**AUTHOR CONTRIBUTIONS**

P.C. designed the project. Q.G., P.C. with the aid of Z.H.T., J.X.P., S.T.C., Y.C.Y, Z.T.L., D.W.S., Z.S., and T.C.C. performed MBE growth, ARPES measurements, and data analysis. Y.H.C. performed calculations. Y.Z.W., and J. J. performed LEED measurements. P.C., Y.H.C., T.C.C. and Q.G. interpreted the results. P.C. wrote the paper with input from other coauthors.

**COMPETING INTERESTS**

Authors declare no competing interests.

**Fig. 1 | Film structure and electronic band structure of single-layer ZrTe₂. a** Atomic structure of single-layer ZrTe₂. **b** Corresponding 2D Brillouin zone. **c** A RHEED pattern of single-layer

ZrTe$_2$ film taken at room temperature. **d** Core-level photoemission spectra taken with 100 eV photons. **e** Schematic diagram for evolution of the band structure and the opening of a gap from a semimetal during the exciton condensation. **f** ARPES maps along $\overline{\Gamma\text{M}}$ taken with $s$ polarized light for the normal phase at 300 K and the condensed phase at 10 K.

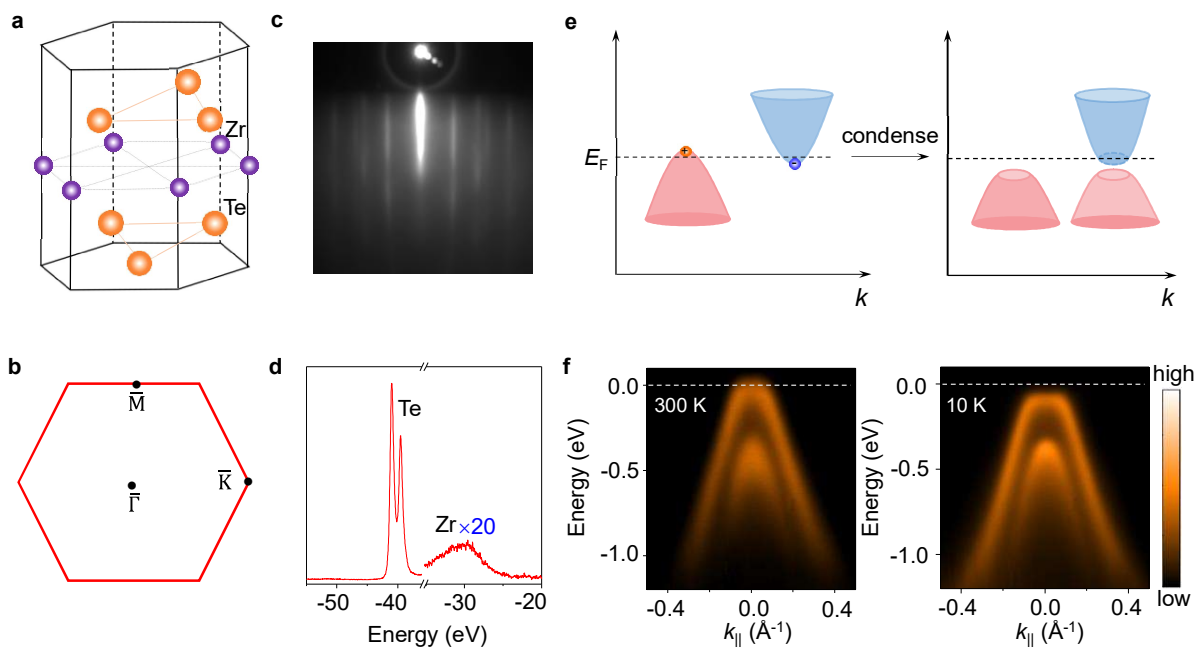

**Fig. 2 | Condensation induced band gap opening and calculated spectra function with exciton order. a** ARPES spectra taken with *p* polarized light along $\overline{\Gamma M}$ direction at 300K and 10 K. The data at 10 K shows the gap opening and folded bands around the $\overline{M}$ point. **b** Calculated spectral functions of single-layer ZrTe$_2$ in the normal and condensed phases. The mean-field solutions of the model in the condensed phase are shown as red dashed curves.

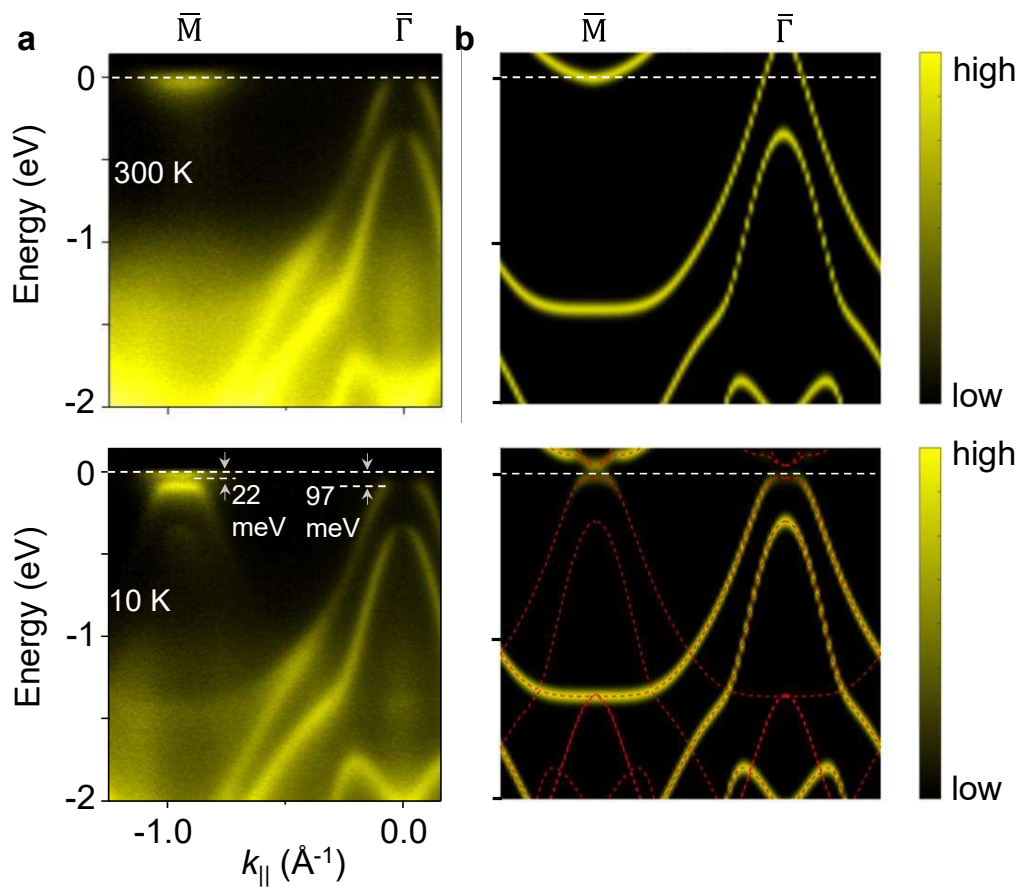

**Fig. 3 | Temperature dependence of the band structure and the Bardeen–Cooper–Schrieffer (BCS)-like behavior of the energy gap**. **a** ARPES spectra along $\overline{\Gamma M}$ reveal that development of the flat valence band top and it shifts away from the Fermi level when the temperature is decreased from 300 to 10 K. The data were taken with 40 eV photons. **b** ARPES maps symmetrized in energy about the Fermi level show a gap in the condensed phase. **c** Symmetrized EDCs at the zone center at selected temperatures between 10 and 300 K. By symmetrization, the effect of Fermi-Dirac distribution at high temperatures can be canceled out. An example of the fit to a phenomenological BCS-type function[34] is shown as a red dashed curve for the EDC obtained at 10 K. **d** The extracted temperature dependence of the square of the energy gap. The red curve is a fit using a BCS-type mean-field equation. Transition temperature $T_C$ is labeled. The error bar is deduced from the standard deviation of the fitting.

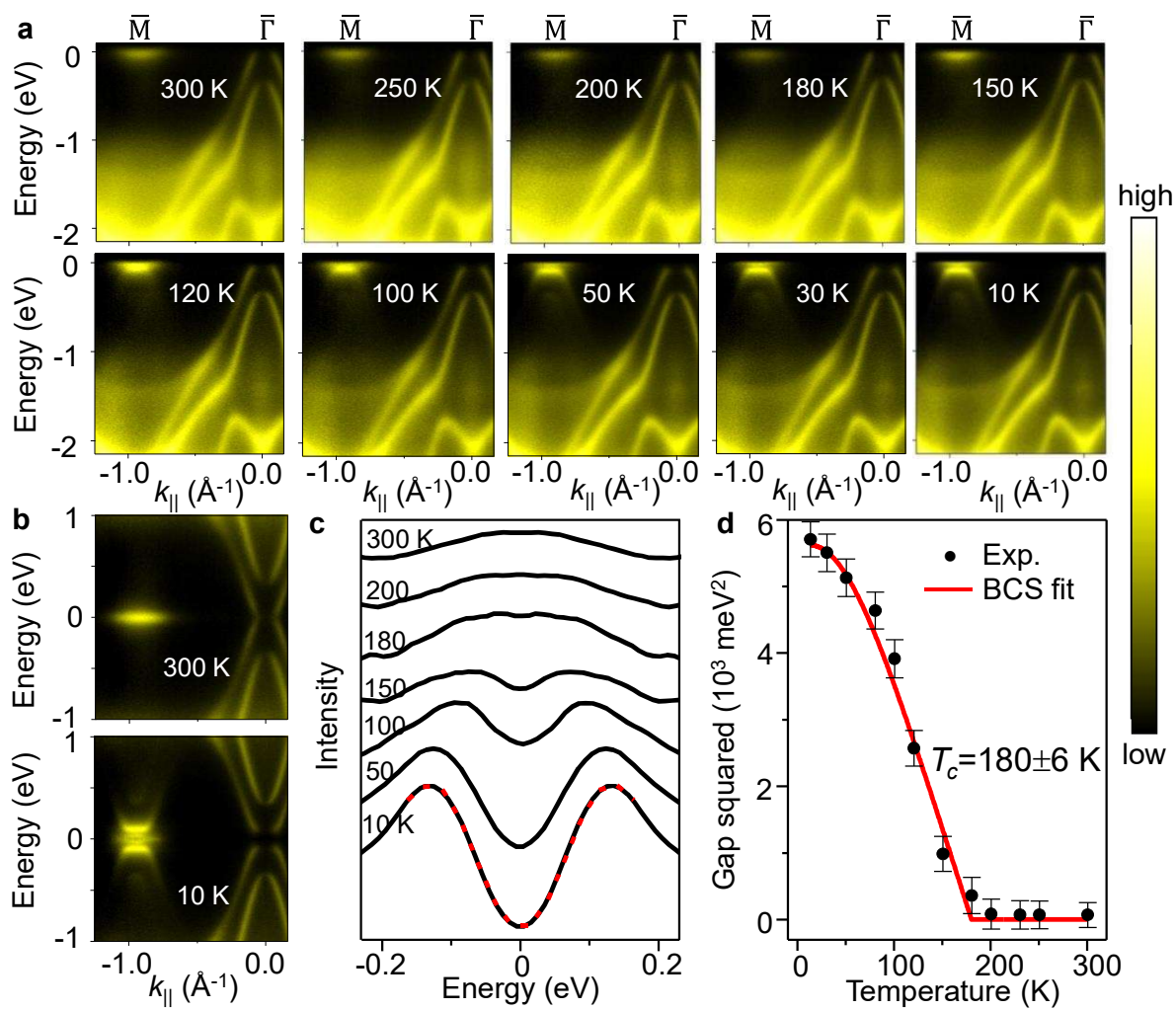

**Fig. 4 | Thickness dependence of the band structure.** ARPES maps along $\overline{\Gamma M}$ for *N*-layer ZrTe$_2$ (*N* = 1, 2, 2.5) taken with unpolarized light (21.2 eV) at **a** 300 K and **b** 10 K. **c** Calculated band dispersions (*N* = 1, 2, and 3) for the normal phase.

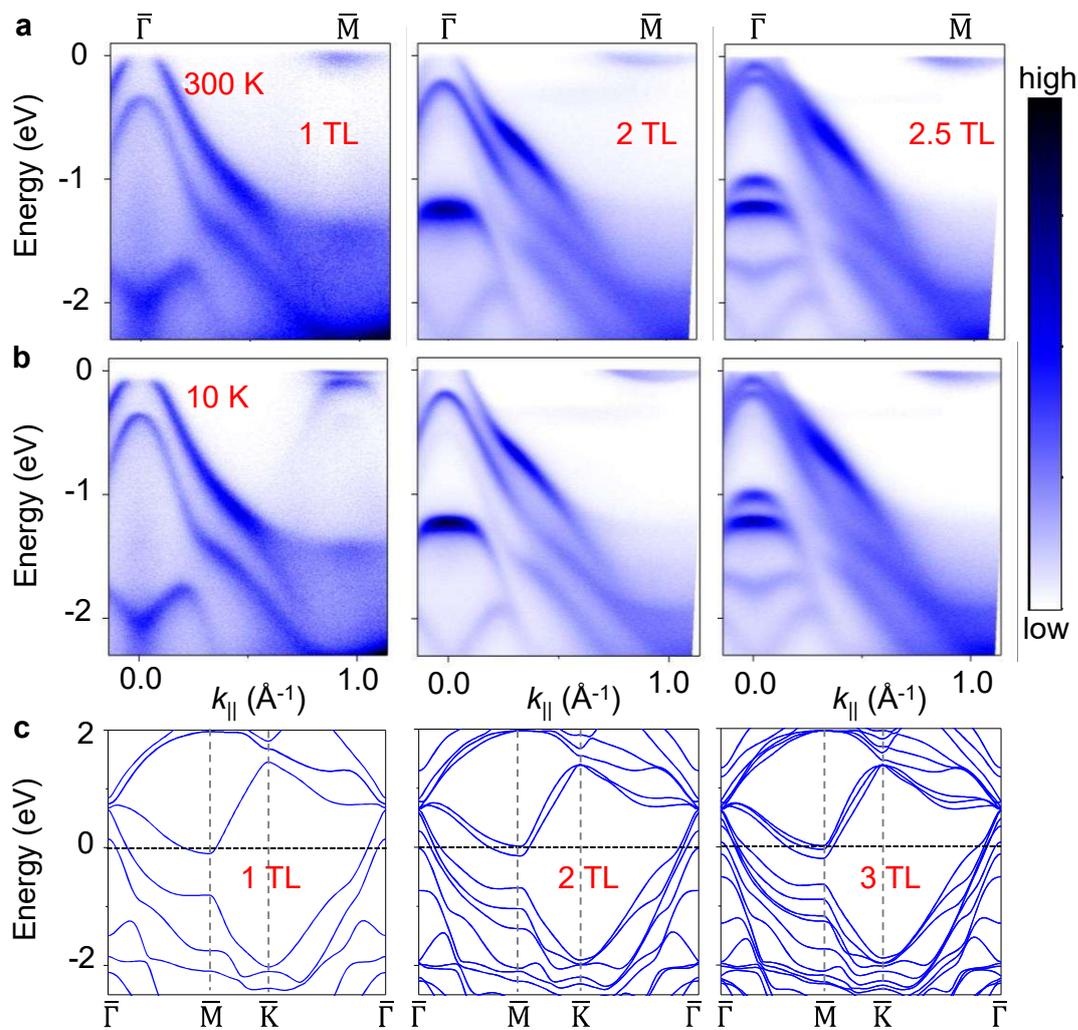